\documentclass[superscriptaddress,aps,prd,aps]{revtex4}
\usepackage{amsmath}
\usepackage{amssymb}
\usepackage{amsfonts}
\usepackage{graphicx,bm}
\usepackage{dcolumn}
\usepackage[colorlinks=true]{hyperref}
\usepackage{epsf}
\usepackage{enumerate}
\usepackage{hhline}
\usepackage{array}
\usepackage{tabularx}
\usepackage{subfigure}

\newcommand{\be}{\begin{equation}}
\newcommand{\ee}{\end{equation}}
\newcommand{\bea}{\begin{eqnarray}}
\newcommand{\eea}{\end{eqnarray}}
\newcommand{\beaa}{\begin{eqnarray*}}
\newcommand{\eeaa}{\end{eqnarray*}}

\newcommand{\nn}{\nonumber \\}

\def\be{\begin{equation}}
\def\ee{\end{equation}}
\def\bea{\begin{eqnarray}}
\def\eea{\end{eqnarray}}

\begin{document}
\title{General slow-roll inflation in $f(R)$ gravity under the Palatini approach}
\author{Sabit Bekov}
\affiliation{Eurasian International Center for Theoretical Physics, Nur-Sultan, 010009, Kazakhstan}
\affiliation{Eurasian  National University, Nur-Sultan, 010008, Kazakhstan}

\author{Kairat Myrzakulov}
\affiliation{Eurasian International Center for Theoretical Physics, Nur-Sultan, 010009, Kazakhstan}
\affiliation{Eurasian  National University, Nur-Sultan, 010008, Kazakhstan}

\author{Ratbay Myrzakulov} 
\affiliation{Eurasian International Center for Theoretical Physics, Nur-Sultan, 010009, Kazakhstan}
\affiliation{Eurasian  National University, Nur-Sultan, 010008, Kazakhstan}

\author{Diego S\'aez-Chill\'on G\'omez} \email{diego.saez@uva.es}
\affiliation{Eurasian International Center for Theoretical Physics, Nur-Sultan, 010009, Kazakhstan}
\affiliation{Department of Theoretical Physics, Atomic and Optics, Campus Miguel Delibes, \\ University of Valladolid UVA, Paseo Bel\'en, 7,
47011 - Valladolid, Spain}

\begin{abstract}
Slow-roll inflation is analyzed in the context of modified gravity within the Palatini formalism. As shown in the literature, inflation in this framework requires the presence of non-traceless matter; otherwise, it does not occur just as a consequence of the nonlinear gravitational terms of the action. Nevertheless, by including a single scalar field that plays the role of the inflaton, slow-roll inflation can be performed in these theories, where the equations lead to an effective potential that modifies the dynamics. We obtain the general slow-roll parameters and analyze a simple model to illustrate the differences introduced by the gravitational terms under the Palatini approach, and the modifications on the spectral index and the tensor to scalar ratio predicted by the model.
\end{abstract}
%
%
\maketitle
%
%
%
\section{Introduction}
General Relativity and other theories constructed from scalar invariants of the Riemann and Ricci tensors assume a metric compatible connection, the so-called Levi--Civita connection, which states that $\nabla_{\lambda}g_{\mu\nu}=0$ and the gravitational action is constructed in terms of such connection and its derivatives. Nevertheless, one may consider both objects in principle as independent fields at the level of the action and then, applying the variational principle to obtain the field equations for both, this is the so-called  Palatini approach. In particular, when a nonlinear function of the Ricci scalar is considered, where the Ricci tensor is constructed in terms of an arbitrary connection, the theory does not lead to General Relativity or the usual metric $f(R)$ equations, but to different ones with diverse phenomenology (for a review, see \citep{Clifton:2011jh,Nojiri:2017ncd,Sotiriou:2006qn,Olmo:2011uz,Vitagliano:2010sr,Capozziello:2011et,Capozziello:2015lza,BeltranJimenez:2018vdo}).

The so-called Palatini $f(\mathcal{R})$ gravity, where $\mathcal{R}=g^{\mu\nu}\mathcal{R}_{\mu\nu}(\Gamma)$ with $\Gamma$ a independent field, has~been widely analyzed in the literature. Specifically, Palatini gravity have been studied in the context of dark energy models and late-time constraints \cite{Baghram:2009we,Aoki:2018lwx,Leanizbarrutia:2017xyd,Boehmer:2013oxa,Capozziello:2012ny,Harko:2011nh,Gu:2018lub,Rosa:2019ejh,Rosa:2017jld,Borowiec:2011wd}, where some models show a very promising behavior as they reproduce the cosmological evolution well and fit the observational constraints correctly~\mbox{\citep{Yang:2008wu,Fay:2007gg,Lee:2007nh}}. Moreover, the modifications on the energy conditions have been also studied in \citep{Capozziello:2014bqa}, and the possible violations of the local gravity tests have been analyzed with the post-Newtonian approach applied to this type of theories \cite{Toniato:2019rrd}, as well as the stellar structure \cite{Olmo:2019flu} and the junction conditions \cite{Olmo:2020fri}, where~significant and interesting differences are found. In a more analytic context, the variational principle \cite{Goenner:2010tr}, the Cauchy problem \cite{Capozziello:2010ut} and Birkhoff's theorem have been explored \cite{Faraoni:2010rt}; in addition, a possible mapping among solutions in GR and these theories are given in \cite{Afonso:2018bpv}. In addition, non-singular black hole solutions seem to be very common in the context of the Palatini formalism \cite{Olmo:2015axa,Olmo:2012nx,Menchon:2017qed} as well as regular wormholes solutions \cite{Bambi:2015zch,Capozziello:2012hr}. In addition, the Palatini formalism is usual in the context of Born--Infeld actions \cite{BeltranJimenez:2017doy}. Moreover, the scalar-tensor picture of Palatini $F(R)$ gravity and its analysis of the Jordan frame and the Einstein frame have been also studied in \citep{Kozak:2018vlp}.

On the other hand, the inflationary paradigm that solves some of the most important theoretical problems of the Big Bang model, is still being deeply analyzed, with many models that reproduce such initial super-accelerating phase \cite{reviews,reviews2,reviews3}. In addition, inflation also produces the required fluctuations for the perturbations that form the seeds on the variation of the matter distribution that formed the galaxies and clusters of galaxies as well as the anisotropies in the Cosmic Microwave Background (CMB) \cite{Mukhanov:1990me,Liddle:1999mq,Langlois:2010xc}. Most of the inflationary models are constructed in terms of scalar fields with the appropriate potential, which generally provides a slow-roll behavior of the scalar field that leads to an exponential expansion at the beginning of inflation and then the field rolls down, leading finally to the reheating regime at the end of inflation \citep{Lidsey:1995np,Elizalde:2008yf}. These models are constructed in such a way that they can satisfy the constraints coming from the analysis of the data collected by Planck over the last few years \citep{Ade:2015lrj,Akrami:2018odb}. Nevertheless, inflation has also been well studied within other underlying models, as in the so-called $f(R)$ gravities \cite{Cognola:2007zu,Nojiri:2003ft,unifying,Cognola:2008zp,BambaOdiDie,delaCruz-Dombriz:2016bjj,Odintsov:2017qif,Sebastiani:2013eqa,Sebastiani:2015kfa,Myrzakulov:2014hca,Bamba:2014jia}, where the $R^2$ model or Starobinsky model \cite{Starobinsky:1980te} shows one of the best behaviors when comparing to the observational data.

The present paper is aimed to analyze the slow-roll scenario in the framework of $f(\mathcal{R})$ gravity under the Palatini approach. Inflation is an scenario that has already been widely analyzed in the literature in the context of Palatini formalism (see \citep{Shimada:2018lnm,ShiMaeda,Gialamas:2020snr,Tenkanen:2020dge,Das:2020kff,Jarv:2017azx,Tenkanen:2017jih,Markkanen:2017tun,Antoniadis:2018yfq,Antoniadis:2018ywb,Enckell:2018hmo,Edery:2019txq,Stachowski:2016zio,Jarv:2020qqm,Koivisto:2005yc,Tamanini:2010uq,Fu:2017iqg,Gialamas:2019nly,Rubio:2019ypq,Rasanen:2018ihz,Rasanen:2018Gu,Gialamas:2020vto,Racioppi:2017spw}). In particular, usual inflaton potentials as the quadratic potential have been considered together with a modification of the Einstein--Hilbert action within the Palatini formalism \citep{Tenkanen:2017jih,Markkanen:2017tun,Antoniadis:2018yfq}, showing better fits to Planck data than single inflaton models with such potentials in GR. In addition, a special emphasis has been done in extending the Starobinsky model of metric $f(R)$ gravity to the Palatini formalism, where, besides the proper gravitational terms in the action, an inflaton scalar field is considered and its modified properties are analyzed in the Einstein frame \citep{Antoniadis:2018ywb,Enckell:2018hmo,Edery:2019txq,Stachowski:2016zio}. One has to note that most of the models studied within the Palatini formalism are considered in the Einstein frame, while the analysis and comparison among frames is well known in this formalism \cite{Kozak:2018vlp}. Moreover, the comparison of inflationary models in Palatini and metric theories in the Einstein frame seems to lead to the same observables \cite{Jarv:2020qqm}. In addition, cosmological perturbation theory has been successfully extended to this framework \citep{Koivisto:2005yc,Tamanini:2010uq,Fu:2017iqg}. In addition, reheating within this $\mathcal{R}^2$ models has also been studied \citep{Gialamas:2019nly,Rubio:2019ypq}. Moreover, some particular models such as Higgs inflation \citep{Rasanen:2018ihz,Rasanen:2018Gu,Gialamas:2020vto} and Coleman--Weinberg inflation \citep{Racioppi:2017spw} have been considered. Here, we aim to obtain the general slow-roll condition with an arbitrary $f(\mathcal{R})$ action in the Palatini formalism with the presence of a scalar field that plays the role of the inflaton that slow rolls, i.e., its kinetic term is very small in comparison to its potential at the beginning of inflation. This analysis is performed completely in the Jordan frame, where we find that the auxiliary scalar field associated with the gravitational part of the action, and~which does not propagate and introduces modifications on the potential of the inflaton, leading to an effective potential that modifies the slow-roll parameters and consequently the spectral index and the tensor to scalar ratio. Then, we assume a quadratic potential for the inflaton and a particular form of the action in order to illustrate the formalism and the differences with respect to General Relativity.

The paper is organized as follows: in Section {\ref{backgroud}, we introduce the Palatini formalism in $f(\mathcal{R})$ gravity theories. Section \ref{InflationFR} reviews the main features of inflation in metric theories and the impossibility of extending such features to the Palatini formalism in vacuum. Then, in Section \ref{SlowRollsect}, we obtain the general slow-roll parameters in these theories with the presence of a scalar field that plays the role of the inflaton. Finally, in Section \ref{conclusions}, we present the conclusions of the paper. 

\section{$f(R)$ Palatini Gravity}
\label{backgroud}
Let us start by reviewing the main equations and features of $f(R)$ gravity a la Palatini. The general gravitational action is given by:
\be
S=\frac{1}{2\kappa^2}\int d^4x \sqrt{-g} f(\mathcal{R}) +S_m \ ,
\label{fRaction}
\ee
where the matter action $S_m$ just depends on the metric and the matter fields, preserving the Equivalence Principle, whereas the Ricci scalar $\mathcal{R}$ is defined as follows:
\be
\mathcal{R}=g^{\mu\nu}\mathcal{R}_{\mu\nu}(\Gamma)=g^{\mu\nu}\left(\partial_{\lambda}\Gamma^{\lambda}_{\mu\nu}-\partial_{\nu}\Gamma^{\lambda}_{\mu\lambda}+\Gamma^{\lambda}_{\sigma\lambda}\Gamma^{\sigma}_{\mu\nu}-\Gamma^{\lambda}_{\sigma\nu}\Gamma^{\sigma}_{\mu\lambda}\right)\ . 
\label{Ricci}
\ee

Here, $\Gamma$ is the connection and in principle is an independent field from the metric, such that, in order to find the field equations, the action (\ref{fRaction}) should be varied with respect to the metric and to the connection, leading to the following field equations respectively:
\bea
  f_{\mathcal{R}}\mathcal{R}_{\mu\nu}-\frac{1}{2}g_{\mu\nu}f=\kappa^2 T_{\mu\nu}\ ,\nn
  \nabla_{\lambda}\left(\sqrt{-g}f_{\mathcal{R}}g^{\mu\nu}\right)=0\ ,
  \label{Fieldeqs}
  \eea
where $T_{\mu\nu}=-\frac{2}{\sqrt{-g}}\frac{\delta S_m}{\delta g^{\mu\nu}}$ is the energy-momentum tensor that only depends on the matter fields and the metric, while $f_{\mathcal{R}}=\frac{df}{d\mathcal{R}}$. Note also that we have not assumed any symmetry on the indexes of the connection, so this might own an antisymmetric part. However, by the projective invariance of the scalar curvature, the Ricci scalar itself only depends on the symmetric part of the connection $R(\Gamma)$ and consequently $\Gamma$ is symmetric under the low indexes. By taking the trace of the first equation in (\ref{Fieldeqs}), it~yields:
\be
  f_{\mathcal{R}}\mathcal{R}-2f=\kappa^2 T\ ,
   \label{traceRT}
   \ee

Hence, this equation provides an algebraic equation for the scalar curvature $\mathcal{R}$ that can be solved as a function of the trace of the energy-momentum tensor $\mathcal{R}=\mathcal{R}(T)$. In addition, the second equation in (\ref{Fieldeqs}) is the metricity condition for the conformal metric:
\be
h_{\mu\nu}=\Omega^2 g_{\mu\nu}\ , \quad \Omega^2=f_{\mathcal{R}}\ ,
\label{conformalTrans}
\ee

such that the second equation in (\ref{Fieldeqs}) results in:
\be
\nabla_{\lambda}\left(\sqrt{-h}h^{\mu\nu}\right)=0\ .
\label{hmetricity}
\ee

As pointed out above, the solution of this equation leads to a connection that is given by the Christoffel symbols in terms of the metric $h_{\mu\nu}$. Then, by applying the conformal transformation (\ref{conformalTrans}), the Ricci tensor $\mathcal{R}_{\mu\nu}$ becomes:
\be
\mathcal{R}_{\mu\nu}(h)=R_{\mu\nu}(g)+\frac{4}{\Omega^2}\nabla_{\mu}\Omega\nabla_{\nu}\Omega-\frac{2}{\Omega}\nabla_{\mu}\nabla_{\nu}\Omega-g_{\mu\nu}\frac{g^{\rho\sigma}}{\Omega^2}\nabla_{\rho}\Omega\nabla_{\sigma}\Omega-g_{\mu\nu}\frac{\Box\Omega}{\Omega}\ .
\label{ConformalR}
\ee

Here, the covariant derivatives are defined in terms of the Levi--Civita connection of the metric $g$. Hence, the first field equation in (\ref{Fieldeqs}) can be written as follows:
	\begingroup\makeatletter\def\f@size{8}\check@mathfonts
\def\maketag@@@#1{\hbox{\m@th\large\normalfont#1}}%
\be
R_{\mu\nu}(g)-\frac{1}{2}g_{\mu\nu}R(g)=\frac{\kappa^2}{f_{\mathcal{R}}}T_{\mu\nu}-g_{\mu\nu}\frac{\mathcal{R}f_{\mathcal{R}}-f}{2f_{\mathcal{R}}}-\frac{3}{2f_{\mathcal{R}}^2}\left[\nabla_{\mu}f_{\mathcal{R}}\nabla_{\nu}f_{\mathcal{R}}-\frac{1}{2}g_{\nu\mu}\nabla_{\lambda}f_{\mathcal{R}}\nabla^{\lambda}f_{\mathcal{R}}\right]+\frac{1}{f_{\mathcal{R}}}\left[\nabla_{\mu}\nabla_{\nu}f_{\mathcal{R}}-g_{\mu\nu}\Box f_{\mathcal{R}}\right]\ .
\label{fieldEq2}
\ee
\endgroup

This equation together with the trace Equation (\ref{traceRT}) form the set of field equations for $f(\mathcal{R})$ in the Palatini formalism that can be applied to study any spacetime in a simple way. Note that the right-hand side of (\ref{fieldEq2}) depends solely on the energy-momentum tensor and its trace, since $\mathcal{R}=\mathcal{R}(T)$ by the Equation (\ref{traceRT}). Moreover, the field Equations (\ref{fieldEq2}) are equivalent to a scalar-tensor theory described by the~action:
\be
S=\frac{1}{2\kappa^2}\int d^4x\sqrt{-g}\left[\phi R(g)+\frac{3}{2\phi}\partial_{\mu}\phi\partial^{\mu}\phi-V(\phi)\right]+S_m\ ,
\label{Brans--DickeEquiv}
\ee
where the correspondence is given by:
\be
 \phi=f_{\mathcal{R}}\ , \quad  V(\phi)=\mathcal{R}\phi-f(\mathcal{R})\ . 
\label{ScalarCorresponde}
\ee 
 
 The Lagrangian (\ref{Brans--DickeEquiv}) is the action for Brans--Dicke theory with a potential and with the usual parameter given by $w=-3/2$. Note that the trace of the field equations in Brans--Dicke theory (with a potential) leads to the dynamical field equation for the scalar field :
 \be
 (3+2w)\Box \phi +2 V(\phi)-\phi\frac{dV}{d\phi}=\kappa^2 T\ ,
 \ee
 which for the Palatini action (\ref{Brans--DickeEquiv}) with $w=-3/2$ becomes:
 \be
 2 V(\phi)-\phi\frac{dV}{d\phi}=\kappa^2 T\ ,
 \label{BDPalatiniAB}
 \ee

Hence, the absence of $\Box \phi$ in the scalar field equation makes the scalar field be non-dynamical, such that the relation among matter and $\phi$ might not interpreted as a strong coupling in the sense as it is in Brans--Dicke theory due to the absence of the excitation induced on $\Box \phi$ by the trace of the energy-momentum tensor. Hence, this equation states an algebraic relation among matter sources and the Palatini action through the auxiliary scalar field $\phi$.  

\section{Inflation in $f(R)$ Gravities}
\label{InflationFR}
Let us first review the main achievements in metric $f(R)$ gravity when describes the inflationary paradigm. In this sense, the so-called Starobinsky inflation \cite{Starobinsky:1980te}} represents the main success of implementing inflation within the modified gravity framework. The model is based on the presence of a quadratic term of the Ricci scalar in the action, which is usually given by:
\be
S=\frac{1}{2\kappa^2}\int {\rm d}^4x\sqrt{-g}\left[R + \frac{R^2}{6\, m^2}\right]\ .
\label{StaroAction}
\ee

Recall that the Ricci scalar in (\ref{StaroAction}) depends on the Levi--Civita connection and consequently the only fundamental field here is the metric tensor. As under the Palatini approach, this action can be expressed in terms of a scalar field:
\be
S=\frac{1}{2\kappa^2}\int {\rm d}^4x\sqrt{-g}\left[\phi R-V(\phi)\right]\ .
\label{BDActionmetricF}
\ee

Here, as above, $\phi=f_{\mathcal{R}}$ and  $V(\phi)=R\phi-f(R)$. However, the action (\ref{BDActionmetricF}) has no kinetic term for the the scalar field, in comparison to (\ref{Brans--DickeEquiv}). Nevertheless, this does not avoid the scalar field in (\ref{BDActionmetricF}) to propagate, as shown by the trace of the field equations for this action, which reads:
\be
 3\Box \phi-R\phi+2V(\phi)=\kappa^2T\ .
 \label{traceMetric}
 \ee

In general, the inflationary model described in metric $f(R)$ gravities is analyzed in the Einstein frame instead of studying the action (\ref{BDActionmetricF}) directly. By performing the conformal transformation:
\be
g_{\mu\nu}\quad \longrightarrow \quad \Omega^2 g_{\mu\nu}\ , \text{where} \quad \Omega^2=f_{R}\ ,
\label{ConformalTrans}
\ee
the action (\ref{BDActionmetricF}) becomes:
\be
\tilde{S}=\int {\rm d}^4x\sqrt{-g_E}\left[\frac{R_E}{2\kappa^2}-\frac{1}{2}\partial_{\mu}\phi_E\partial^{\mu}\phi_E-V_E(\phi_E)\right]\ ,
\label{EinsteinFrame}
\ee
where the subindex $_E$ refers to the quantities transformed to the Einstein frame. The potential $V_E(\phi_E)$ for the Starobinsky action (\ref{StaroAction}) turns out:
\be
V_E( \phi_E) = \frac{3}{4 \kappa^2} m^2 \left( 1 - e^{- \sqrt{2/3} \kappa \phi_E} \right)^2 .
\label{VEinstein}
\ee

We assume a flat FLRW metric in the Einstein frame:
\be
ds^2=-dt^2+a(t)^2\sum_i dx_i^2\ .
\label{FLRWmetric}
\ee

The corresponding FLRW equations lead to:
\be
H^2=\frac{\kappa^2}{3}\left(\frac{1}{2}\dot{\phi}_E^2+V_E(\phi_E)\right)\ , \quad \dot{H}=-\frac{\kappa^2}{2}\dot{\phi}_E^2\ ,
\label{FLRWeqsEins}
\ee
while the equation for the scalar field $\phi$ is obtained by varying the action (\ref{EinsteinFrame}) with respect to the scalar~field:
\be
\ddot{\phi}_E+3H\dot{\phi}_E+V'(\phi_E)=0\ .
\label{ScalarFieldEqEins}
\ee

Hence, the FLRW Equations (\ref{FLRWeqsEins}) together with the scalar field Equation (\ref{ScalarFieldEqEins}) may correspond to inflationary models with a single scalar field minimally coupled. The slow-roll scenario occurs in the regime $\frac{1}{2}\dot{\phi}_E^2<<V_E(\phi_E)$ and $\ddot{\phi}_E<<H\dot{\phi}_E$, which approximates the first FLRW equation and the scalar field equation to:
\be
H^2\approx\frac{\kappa^2}{3}V_E(\phi_E)\ , \quad 3H\dot{\phi}_E+V'(\phi_E)\approx0\ .
\label{FLRWeqsEins}
\ee

This is the slow-roll approximation, which is equivalent to the following conditions on the so-called slow-roll parameters:
\be
\epsilon=-\frac{\dot{H}}{H^2}\approx \frac{1}{2\kappa^2}\left(\frac{V_E'}{V}\right)^2<<1\ , \quad |\eta|\approx \frac{1}{\kappa^2}\left(\frac{V_E''}{V_E}\right)<<1\ .
\label{SlowRollparam}
\ee

Hence, for the appropriate scalar potential, the slow-roll inflationary scenario can be implemented. Actually, this is the case for the potential (\ref{VEinstein}), whose slow-roll parameters become:
\bea
\epsilon &\simeq& \frac{4}{3} \left(e^{\sqrt{\frac{2}{3}}\kappa \phi_E} -1 \right)^{-2}  \simeq \frac{4}{3} e^{-2\sqrt{\frac{2}{3}}\kappa \phi_E}\ , \\ 
\eta &\simeq&  \frac{4}{3} \frac{2-e^{\sqrt{\frac{2}{3}}\kappa \phi_E}}{\left(-1+e^{\sqrt{\frac{2}{3}}\kappa \phi_E}\right)^{2}}  \simeq -\frac{4}{3} e^{-\sqrt{\frac{2}{3}}\kappa \phi_E}.
\eea

Here, we have assumed that the scalar field is large enough at the beginning of inflation $\kappa\phi_E>>1$, which makes $\epsilon<<1$ and $\eta<<1$, and the Hubble parameter is approximately a constant, which leads to an exponential expansion. Inflation ends when the scalar field rolls down and its kinetic term becomes larger, which results in $\epsilon\approx1$. The important point of inflation is to last enough in order to solve the initial conditions problems, such that it should last a sufficient number of e-foldings:
\be
N  = \int_{t_{start}}^{t_{end}} H dt\approx 55-65\ .
\label{N_Staro}
\ee

By using the FLRW equation, this can be computed in terms of the scalar field:
\be
N = -\kappa^2 \int_{\phi_{Ei}}^{\phi_{Eend}} \frac{V_E(\phi_E}{V_E'(\phi_E)} {\rm d}\phi_E\approx\frac{3}{4}e^{\sqrt{2/3}\kappa \; \phi_{Ei}}\ .
\label{Nefolds}
\ee

Note that the number of e-foldings is related to the slow-roll parameters as follows:
\be
\epsilon \simeq \frac{3}{4}\frac{1}{N^2}\ , \quad \eta \simeq -\frac{1}{N}\ .
\label{SRparamStaro}
\ee

While the spectral index for curvature perturbations and the tensor to scalar ratio are given by:
\be
n_s-1=-6\epsilon+2\eta\ , \quad r=16\epsilon
\label{nsr}
\ee

The last constraints on both parameters from Planck \cite{Ade:2015lrj,Akrami:2018odb} are:
\be
n_s=0.9659\pm 0.0041\ , \quad r<0.11\ .
\label{constraintsPlanck1}
\ee

Hence, for the Starobinsky inflation by assuming $N=65$, we obtain:
\be
n_s=0.968\ , \quad r=0.00284\,
\label{SpRStaro}
\ee
which fit the above constraints very well and this is one of the great success of Starobinsky inflation and metric $f(R)$ inflation in general. 

Nevertheless, as shown in the previous section, this is not the case for Palatini $f(\mathcal{R})$ gravity. From~the trace Equation (\ref{BDPalatiniAB}), one directly notes that, in comparison to the metric case (\ref{traceMetric}), the scalar field does not propagate, and in absence of matter, the scalar field becomes a constant, such that it only provides and effective cosmological constant, which produces an accelerating expansion with no end and obviously with no perturbations. Hence, the only way to produce inflation in Palatini $f(\mathcal{R})$ gravity is in the presence of non-traceless matter, since the presence of a scalar field that plays the role of the inflaton induces modifications on the slow-roll scenario, as is shown in the next section.

\section{Slow-Roll Inflation in Palatini $f(\mathcal{R})$ Gravity}
\label{SlowRollsect}

As pointed above, since the scalar field associated with the $f(\mathcal{R})$ action in the Palatini approach does not propagate and its source is the presence of non traceless matter, the only way to reproduce inflation is to include an inflaton field. Nevertheless, as will be shown below, the slow-roll conditions are modified in comparison to General Relativity when a Palatini $f(\mathcal{R})$ Lagrangian is assumed. Then,~here~we will analyze the Palatini $f(\mathcal{R})$ Lagrangian in its version expressed in terms of an auxiliary scalar field as given in (\ref{Brans--DickeEquiv}), with the presence of an additional scalar field $\chi$ that plays the role of the inflaton:
\be
S=\frac{1}{2\kappa^2}\int d^4x\sqrt{-g}\left[\phi R(g)+\frac{3}{2\phi}\partial_{\mu}\phi\partial^{\mu}\phi-V(\phi)\right]+\int d^4x\sqrt{-g}\left[-\frac{1}{2}\partial_{\mu}\chi\partial^{\mu}\chi-U(\chi)\right]\ .
\label{InflatonPalatini}
\ee

The corresponding field equations are given by (\ref{fieldEq2}), where we recall that $\phi=f_{\mathcal{R}}$ and  $V(\phi)=\mathcal{R}\phi-f(\mathcal{R})$ and the energy-momentum tensor $\chi$:
 \be
 T_{\mu\nu}^{\chi}=\partial_{\mu}\chi\partial_{\nu}\chi-g_{\mu\nu}\left(\frac{1}{2}\partial_{\alpha}\chi\partial^{\alpha}\chi+U(\chi)\right)\ .
 \label{chi_EMtensor}
 \ee

By assuming a flat FLRW metric as given in (\ref{FLRWmetric}), the FLRW equations yield:
\bea
3H^2&=&\frac{\kappa^2}{\phi}\left(\frac{1}{2}\dot{\chi}^2+U(\chi)\right)-3H\frac{\dot{\phi}}{\phi}-\frac{3}{4}\left(\frac{\dot{\phi}}{\phi}\right)^2+\frac{1}{2\phi}V(\phi)\ , \nn
-2\dot{H}&=&\frac{\kappa^2}{\phi}\dot{\chi}^2-H\frac{\dot{\phi}}{\phi}-\frac{3}{2}\left(\frac{\dot{\phi}}{\phi}\right)^2+\frac{\ddot{\phi}}{\phi}\ ,
\label{FLRWeqsphichi}
\eea
while the equations for the scalar fields are given by:
\bea
2V(\phi)-\phi V'(\phi)&=&\kappa^2\left(\dot{\chi}^2-4U(\chi)\right)\ , \nn
\ddot{\chi}+3H\dot{\chi}+U'(\chi)&=&0\ .
\label{PhichiEqs}
\eea

We can now consider the scalar field $\chi$ to be in the slow roll regime during the inflationary phase, as in the usual single field inflation in GR. Then, the following slow-roll conditions are assumed on $\chi$:
\be
U(\chi)>>\frac{1}{2}\dot{\chi}^2\ , \quad H\dot{\chi}>>\ddot{\chi}\ .
\label{SRconditionschi}
\ee

Hence, the scalar field Equation (\ref{PhichiEqs}) reduces to:
\bea
2V(\phi)-\phi V'(\phi)&\approx &-4\kappa^2 U(\chi)\ , \nn
3H\dot{\chi}+U'(\chi)&\approx&0\ .
\label{PhichiEqs2}
\eea

From here, we end to the conclusion that the problem is reduced to a single scalar field inflationary model, since, from the first equation in (\ref{PhichiEqs2}), one gets $\phi\approx\phi(\chi)$, as far as the equation has real solutions in terms of $\phi$. This is not surprising, as Equation (\ref{PhichiEqs}) for $\phi$ does not contain first or second derivatives of the scalar field, an aspect already discussed in the section above, inherent to the Palatini approach that avoids the auxiliary scalar field $\phi$ to propagate. Moreover, as a consequence of (\ref{PhichiEqs2}) during inflation, we can assume that the auxiliary field would behave approximately as a constant, such that $\frac{\dot{\phi}^2}{V(\phi)}<< 1$ and $\frac{\ddot{\phi}}{H\dot{\phi}}<<1$ hold in (\ref{FLRWeqsphichi}). Hence, by using the slow-roll approximations (\ref{SRconditionschi}) and the scalar field Equation (\ref{PhichiEqs2}), the term $H\dot{\phi}$ can be expressed as follows:
\be
H\dot{\phi}=H\frac{d\phi}{d\chi}\dot{\chi}\approx - \frac{1}{3}\frac{d\phi}{d\chi} U'(\chi)\ ,
\ee
while the terms $\dot{\phi}^2$ and $\ddot{\phi}$ can be neglected in the FLRW Equation (\ref{FLRWeqsphichi}), which can be expressed just in terms of the inflaton field $\chi$ and are approximated as:
\bea
3H^2&\approx &\frac{1}{\phi}\left[\kappa^2 U(\chi)+\frac{d\phi}{d\chi}U'(\chi)+\frac{1}{2}V(\phi(\chi))\right]=\kappa^2 U_{eff}(\chi)\ , \nn
-2\dot{H}&\approx &\frac{1}{\phi}\left(\kappa^2\dot{\chi}^2+\frac{1}{3}\frac{d\phi}{d\chi}U'(\chi)\right)\approx \frac{1}{3\phi}\left(\frac{U^{'2}(\chi)}{U_{eff}(\chi)}+\frac{d\phi}{d\chi}U'(\chi)\right)\ .
\label{FLRWeqsphichi2}
\eea

Recall that $\phi=\phi(\chi)$, such that these equations only depend on the dynamics of the inflaton field~$\chi$. In addition, here we have used the slow-roll conditions on $\chi$ (\ref{SRconditionschi}) and the scalar field Equation~(\ref{PhichiEqs2}), while we have defined the effective potential as:
\be
U_{eff}=\frac{1}{\kappa^2\phi}\left[\kappa^2 U+\frac{d\phi}{d\chi}U'(\chi)+\frac{1}{2}V(\phi(\chi))\right]\ .
\label{Uefectivo}
\ee
 
 Hence, the slow-roll parameters (\ref{SlowRollparam}) will read now differently:
 \bea
 \epsilon&\approx & \frac{1}{2\kappa^2\phi}\left[\left(\frac{U'}{U_{eff}}\right)^2+\frac{d\phi}{d\chi}\frac{U'}{U_{eff}}\right]\ ,\nn
 |\eta| &\approx & \frac{1}{\kappa^2}\frac{U''}{U_{eff}}\ .
 \label{SlowrollParamLast}
 \eea

Note that, for $f(\mathcal{R})=\mathcal{R}=R(g)$, i.e., when General Relativity is recovered, the scalar field $\phi=1$ and the usual slow-roll parameters (\ref{SlowRollparam}) are recovered in terms of the inflaton potential $U$. Hence,~slow-roll inflation will occur when $\epsilon<<1$ and $\eta<<1$, while inflation ends for $\dot{\chi}^2\approx 2 U(\chi)$, when the approximation (\ref{PhichiEqs2}) is no longer valid. 

Let us consider a model to illustrate the differences introduced by the modified gravitational action $f(\mathcal{R})$. To do so, we choose the following-commonly used- potential for the inflaton:
\be
U(\chi)=\frac{1}{2}m_{\chi}^2\dot{\chi}^2\ .
\label{InlaftonU}
\ee

This potential has been widely analyzed in the literature and describes the so-called chaotic inflation, which under the slow-roll conditions (\ref{SRconditionschi}) in General Relativity gives:
\be 
H^2\approx \frac{\kappa^2}{6}m_{\chi}^2\chi^2\ .
\label{GRChaoticH}
\ee
while the number of e-foldings (\ref{Nefolds}) can be expressed in terms of the scalar field as follows:
\be
N = -\kappa^2 \int_{\chi_{i}}^{\chi_{end}} \frac{U(\chi)}{U'(\chi)} {\rm d}\chi\approx\frac{\kappa^2\chi_i^2}{4}\ .
\label{NefoldsGR}
\ee

Hence, the slow-roll parameters (\ref{SlowRollparam}) are:
\be
\epsilon=\eta=\frac{1}{N}\ ,
\ee
which, for a number of e-foldings, $N=65$ provides the following values for the spectral index and the tensor to scalar ratio:
\be
n_s=0.97\ , \quad r=0.12\ .
\label{nsrGR}
\ee

By comparing with the last data from Planck (\ref{constraintsPlanck1}), it is clear that the $r$ does not satisfy the~constraint. 

Let us now analyze the same potential for the inflaton (\ref{InlaftonU}) in the context of Palatini gravity when assuming the following gravitational action:
\be
f(\mathcal{R})=\alpha\mathcal{R}^n\ ,
\label{Rninflation}
\ee
where $\alpha$ is a constant with the appropriate dimensions and $n$ is also a constant. Note that this type of gravitational terms has been considered previously in the literature in the framework of Palatini formalism together with a linear term of the Ricci scalar, which show a good behavior in order to reproduce adequately the cosmological evolution \citep{Yang:2008wu,Fay:2007gg} and, particularly, the $R^n$ term shows similar results in comparison to GR when dealing with cosmological perturbations \cite{Lee:2007nh}. In addition, such type of gravitational actions are capable of satisfying the local constraints, as shown by its post-Newtonian approximation \cite{Toniato:2019rrd}. The corresponding auxiliary scalar field $\phi$ and its potential are then provided by~(\ref{ScalarCorresponde}) and lead to:
\be
\phi=n\alpha \mathcal{R}^{n-1}\ , \quad V(\phi)=k\ \phi^{n/n-1}\ ,
\label{ScalarCorresponde2}
\ee
here $k=\frac{1}{(n\alpha)^{1/(n-1)}}\left(\frac{n-1}{n}\right)$. By assuming the slow-roll conditions for the scalar field $\chi$ given in (\ref{SRconditionschi}), the~relation $\phi=\phi(\chi)$ is obtained from the first equation in (\ref{PhichiEqs2}):
\be
\phi(\chi)\approx \left(\frac{k(n-2)}{2m_{\chi}^2(n-1)}\frac{1}{\kappa^2\chi^2}\right)^{\frac{1-n}{n}}\ .
\label{phichiRel}
\ee

From here, we can deduce that $n\neq 1,2$, since, for $n=1$, we recover General Relativity and $n=2$ is a degenerate case as shown by the trace Equation (\ref{traceRT}). The Hubble parameter in (\ref{FLRWeqsphichi2}) leads to:
\be
3 H^2=\frac{\kappa^2 m_{\chi}^2}{2k\ n(n-2)}\frac{2^{1/n}k\ n(3n-4)+8m_{\chi}^2(n-1)^2 \left(\frac{k(n-2)}{(n-1)m_{\chi}^2\kappa^2\chi^2}\right)^{1/n}}{ \left(\frac{k(n-2)}{(n-1)m_{\chi}^2\kappa^2\chi^2}\right)^{\frac{1-n}{n}}}\chi^2\ ,
\label{HubbleExample1}
\ee
which for the case $n>2$ and assuming a large field at the beginning of inflation $\kappa\chi>>1$, it yields:
\be
3H^2\approx \frac{2^{\frac{1-n}{n}}(3n-4)\kappa^2 m_{\chi}^2}{(n-2)}\left(\frac{k(n-2)}{(n-1)m_{\chi}^2\kappa^2\chi^2}\right)^{\frac{|1-n|}{n}}\chi^2\ .
\label{HubbleExample1b}
\ee

In order to simplify the expressions, we can assume, for illustrative purposes, the case $n=3$. Hence, the slow-roll parameters (\ref{SlowrollParamLast}) turn out:
\be
\epsilon=\frac{5\cdot 2^{4/3}m_{\chi}^2}{75 k}\left(\frac{k}{m_{\chi}^2\kappa^2\chi^2}\right)^{1/3}\ , \quad \eta=\frac{2^{7/3}m_{\chi}^2}{5 k}\left(\frac{k}{m_{\chi}^2\kappa^2\chi^2}\right)^{1/3}\ ,
\label{SlowRollExample1}
\ee
while the number of e-foldings is given by:
\be
N=\int_{t_i}^{t_{end}}H dt=-\kappa^2\int_{\chi_i}^{\chi_{end}}\frac{U_{eff}}{U'}d\chi=\frac{15}{8\cdot 2^{1/3}}\left(\frac{k}{m_{\chi}^2}\right)^{2/3}(\kappa^2\chi_i^2)^{1/3}\ .
\label{EfoldsEx1}
\ee

Then, the slow-roll parameters (\ref{SlowRollExample1}) can be expressed in terms of the number of e-foldings as~follows:
\be
\epsilon=\frac{13}{10}\frac{1}{N}\ , \quad \eta=\frac{3}{2}\frac{1}{N}\ ,
\label{SlowRollExample1b}
\ee
while the spectral index and the tensor to scalar ratio lead to:
 \be
n_s=0.92\ , \quad r=0.34\ ,
\label{nsrn3}
\ee
which neither satisfy the constraints from Planck (\ref{constraintsPlanck1}). Actually, any integer for $n>2$ does not provide good fits to Planck data. Nevertheless, for half integers, the predictions are different. We may consider $n=4/5$, in whose case the Hubble parameter yields:
\be
\epsilon=\frac{3^{5/4}m_{\chi}^2}{4 k}\left(\frac{k}{m_{\chi}^2\kappa^2\chi^2}\right)^{5/4}\ , \quad \eta=\frac{3^{5/4}m_{\chi}^2}{2 k}\left(\frac{k}{m_{\chi}^2\kappa^2\chi^2}\right)^{5/4}\ ,
\label{SlowRollExample2}
\ee
whereas the number of e-foldings is given by:
\be
N=\frac{4}{5\cdot 3^{5/4}}\left(\frac{m_{\chi}^2}{k}\right)^{1/4}(\kappa^2\chi_i^2)^{5/4}\ .
\label{EfoldsEx1}
\ee

Hence, the slow-roll parameters in terms of $N$ give:
\be
\epsilon=\frac{1}{5}\frac{1}{N}\ , \quad \eta=\frac{2}{5}\frac{1}{N}\ ,
\label{SlowRollExample2b}
\ee
which, for $N=65$, we obtain the following spectral index and tensor to scalar ratio:
 \be
n_s=0.98\ , \quad r=0.05\ ,
\label{nsrn3}
\ee
which fit the Planck constraints much better, particularly the tensor to scalar ratio. Hence, as illustrated by this simple model of chaotic inflation and a power gravitational action, the corrections introduced in the Palatini formalism play an important role in the viability of a particular inflaton model.

\section{Conclusions}
\label{conclusions}

In the present paper, we have analyzed the inflationary paradigm in the framework of $f(\mathcal{R})$ Palatini gravity. As shown in other previous papers, and as a consequence of the non-propagating scalar field in the Palatini formalism, in vacuum or with the presence of traceless matter, the action $f(\mathcal{R})$ just adds an effective cosmological constant that is not enough to construct a viable model for inflation, since inflation should end after a particular number of e-foldings. Hence, the only way is to consider the construction of inflationary models in this framework is to consider the presence of matter, in general a single scalar field with an appropriate potential, similarly as in General Relativity. 

Nevertheless, the presence of nonlinear functions of the Ricci scalar $\mathcal{R}$ in the gravitational action induces modifications on the dynamics of the Hubble parameter and consequently on the inflaton, leading to an effective potential. Here, by working in the Jordan frame, we have obtained such effective potential and have calculated the modified slow-roll parameters when one assumes a slow-roll motion for the inflaton. As a consequence of these modifications, the corresponding spectral index and tensor to scalar ratio are modified too, such that the same potential for the inflaton does not provide the same predictions in GR and in $f(\mathcal{R})$ Palatini gravity. We have illustrated this feature by a simple case, the so-called chaotic inflation which is described by a quadratic potential for the inflaton. As shown, this model that does not fit the Planck constraints well is slightly modified by a nonlinear term of the Ricci scalar in the action, leading to much better predictions for the spectral index and the tensor to scalar ratio.

Hence, slow-roll inflation in modified gravities within the Palatini approach reflects important differences that may help to satisfy the observational constraints for some models that in the context of GR were ruled out.

\vspace{6pt}


\acknowledgments{DS-CG is acknowledges support from the University of Valladolid (Spain). The work was financially supported by the Ministry of Education and Science of the Republic of Kazakhstan, Grant No. AP08052034.}

%
%

%


\begin{thebibliography}{999}

\bibitem{Clifton:2011jh}
Clifton, T.; Ferreira, P.G.; Padilla, A.; Skordis, C.
 Modified Gravity and Cosmology. 
\emph{Phys. Rept.} \textbf{2012}, \emph{513}, 1--189,
doi:10.1016/j.physrep.2012.01.001; 
\bibitem{Nojiri:2017ncd}
Nojiri, S.; Odintsov, S.D.; Oikonomou,~V.K.
 Modified Gravity Theories on a Nutshell: Inflation, Bounce and Late-time Evolution. 
\emph{Phys. Rept.} \textbf{2017}, \emph{692}, 1--104,
doi:10.1016/j.physrep.2017.06.001; 
\bibitem{Sotiriou:2006qn}
Sotiriou, T.P.; Liberati, S.
 Metric-affine f(R) theories of gravity. 
\emph{Ann. Phys.} \textbf{2007}, \emph{322}, 935--966,
doi:10.1016/j.aop.2006.06.002; 
\bibitem{Olmo:2011uz}
Olmo, G.J.
 Palatini Approach to Modified Gravity: f(R) Theories and Beyond. 
\emph{Int. J. Mod. Phys. D} \textbf{2011}, \emph{20}, 413--462,
doi:10.1142/S0218271811018925; 
\bibitem{Vitagliano:2010sr}
Vitagliano, V.; Sotiriou, T.P.; Liberati, S.
 The dynamics of metric-affine gravity. 
\emph{Ann. Phys.} \textbf{2011}, \emph{326}, 1259--1273,
[erratum: \emph{Ann. Phys.} \textbf{2013}, \emph{329}, 186--187],
doi:10.1016/j.aop.2011.02.008; 
\bibitem{Capozziello:2011et}
Capozziello, S.; Laurentis, M.D.
 Extended Theories of Gravity. 
\emph{Phys. Rep.} \textbf{2011}, \emph{509}, 167--321,
doi:10.1016/j.physrep.2011.09.003; 
\bibitem{Capozziello:2015lza}
Capozziello, S.; Harko, T.; Koivisto, T.S.; Lobo, F.S.N.; Olmo, G.J.
 Hybrid metric-Palatini gravity. 
\emph{Universe} \textbf{2015}, 1, 199--238,
doi:10.3390/universe1020199; 
\bibitem{BeltranJimenez:2018vdo}
Beltr\'an Jim\'enez, J.; ~Heisenberg, L.; Koivisto, T.S.
 Teleparallel Palatini theories. 
\emph{JCAP} \textbf{2018}, \emph{8}, 39,
doi:10.1088/1475-7516/2018/08/039.



\bibitem{Baghram:2009we}
Baghram, S.; Rahvar, S.
 Inverse problem: Reconstruction of modified gravity action in Palatini formalism by Supernova Type Ia data. 
\emph{Phys. Rev. D} \textbf{2009}, \emph{80}, 124049,
doi:10.1103/PhysRevD.80.124049; 
\bibitem{Aoki:2018lwx}
Aoki, K.; Shimada, K. 
 Galileon and generalized Galileon with projective invariance in a metric-affine formalism. 
\emph{Phys. Rev. D} \textbf{2018}, \emph{98}, 044038,
doi:10.1103/PhysRevD.98.044038; 
\bibitem{Leanizbarrutia:2017xyd}
Leanizbarrutia, I.; Lobo, F.S.N.; Saez-Gomez, D.
 Crossing SNe Ia and BAO observational constraints with local ones in hybrid metric-Palatini gravity. 
\emph{Phys. Rev. D} \textbf{2017}, \emph{95}, 084046,
doi:10.1103/PhysRevD.95.084046; 
\bibitem{Boehmer:2013oxa}
B\"ohmer, C.G.; Lobo, F.S.N.; Tamanini, N.
 Einstein static Universe in hybrid metric-Palatini gravity. 
\emph{Phys. Rev. D} \textbf{2013}, \emph{88}, 104019,
doi:10.1103/PhysRevD.88.104019; 
\bibitem{Capozziello:2012ny}
Capozziello,~S.; Harko, T.; Koivisto, T.S.; Lobo, F.S.N.; Olmo, G.J.
 Cosmology of hybrid metric-Palatini f(X)-gravity. 
\emph{JCAP} \textbf{2013}, \emph{4}, 11,
doi:10.1088/1475-7516/2013/04/011; 
\bibitem{Harko:2011nh}
Harko, T.; Koivisto, T.S.; Lobo, F.S.N.; Olmo, G.J.
 Metric-Palatini gravity unifying local constraints and late-time cosmic acceleration. 
\emph{Phys. Rev. D} \textbf{2012}, \emph{85}, 084016,
doi:10.1103/PhysRevD.85.084016; 
\bibitem{Gu:2018lub}
Gu, B.M.; Liu, Y.X.; Zhong, Y.
 Stable Palatini $f(\mathcal{R})$ braneworld. 
\emph{Phys. Rev. D} \textbf{2018}, \emph{98}, 024027,
doi:10.1103/PhysRevD.98.024027; 
\bibitem{Rosa:2019ejh}
Rosa, J.L.; Carloni, S.; Lemos,~J.P.S.
 Cosmological phase space of generalized hybrid metric-Palatini theories of gravity. 
\emph{Phys. Rev. D} \textbf{2020}, \emph{1}, 104056,
doi:10.1103/PhysRevD.101.104056; 
\bibitem{Rosa:2017jld}
Rosa, J.L.; Carloni, S.; Lemos, J.P.d.; Lobo, F.S.N. 
 Cosmological solutions in generalized hybrid metric-Palatini gravity. 
\emph{Phys. Rev. D} \textbf{2017}, \emph{95}, 124035,
doi:10.1103/PhysRevD.95.124035; 
\bibitem{Borowiec:2011wd}
Borowiec, A.; Kamionka, M.; Kurek, A.; Szydlowski, M.
 Cosmic acceleration from modified gravity with Palatini formalism. 
\emph{JCAP} \textbf{2012}, \emph{2}, 27,
doi:10.1088/1475-7516/2012/02/027. 


\bibitem{Yang:2008wu}
Yang, X.J.; Chen, D.M.
 $f(R)$ gravity theories in the Palatini Formalism constrained from strong lensing. 
\emph{Mon. Not. Roy. Astron. Soc.} \textbf{2009}, \emph{394}, 1449,
doi:10.1111/j.1365-2966.2008.14318.x.


\bibitem{Fay:2007gg}
Fay, S.; Tavakol, R.; Tsujikawa, S.
 f(R) gravity theories in Palatini formalism: Cosmological dynamics and observational constraints. 
\emph{Phys. Rev. D} \textbf{2007}, \emph{75}, 063509,
doi:10.1103/PhysRevD.75.063509.

\bibitem{Lee:2007nh}
Lee, S.
Stability of Palatini-f(R) cosmology.  \emph{arXiv}  \textbf{2007}, arXiv:0710.2395 [gr-qc].

\bibitem{Capozziello:2014bqa}
Capozziello, S.; Lobo, F.S.N.; Mimoso, J.P.
 Generalized energy conditions in Extended Theories of Gravity. 
\emph{Phys. Rev. D} \textbf{2015}, \emph{91}, 124019,
doi:10.1103/PhysRevD.91.124019.


\bibitem{Toniato:2019rrd}
Toniato, J.D.; Rodrigues, D.C.; Wojnar, A.
 Palatini $f(R)$ gravity in the solar system: post-Newtonian equations of motion and complete PPN parameters. 
\emph{Phys. Rev. D} \textbf{2020}, \emph{101}, 064050,
doi:10.1103/PhysRevD.101.064050.

\bibitem{Olmo:2019flu}
Olmo, G.J.; Rubiera-Garcia, D.; Wojnar, A.
 Stellar structure models in modified theories of gravity: lessons and challenges. 
\emph{Phys. Rep.} \textbf{2020}, \emph{876}, 1--75,
doi:10.1016/j.physrep.2020.07.001.

\bibitem{Olmo:2020fri}
Olmo, G.J.; Rubiera-Garcia, D.
 Junction conditions in Palatini $f(R)$ gravity. 
\emph{Class. Quant. Grav.} \textbf{2020}, \emph{37}, 215002,
doi:10.1088/1361-6382/abb924.

\bibitem{Goenner:2010tr}
Goenner, H.F.M.
 Alternative to the Palatini method: A new variational principle. 
\emph{Phys. Rev. D} \textbf{2010}, \emph{81}, 124019,
doi:10.1103/PhysRevD.81.124019.

\bibitem{Capozziello:2010ut}
Capozziello, S.; Vignolo, S.
 The Cauchy problem for metric-affine f(R)-gravity in the presence of a Klein-Gordon scalar field. 
\emph{Int. J. Geom. Meth. Mod. Phys.} \textbf{2011}, \emph{8}, 167--176,
doi:10.1142/S0219887811005063.

\bibitem{Faraoni:2010rt}
Faraoni, V.
 The Jebsen-Birkhoff theorem in alternative gravity. 
\emph{Phys. Rev. D} \textbf{2010}, \emph{81}, 044002,
doi:10.1103/PhysRevD.81.044002.


\bibitem{Afonso:2018bpv}
Afonso, V.I.; Olmo, G.J.; Rubiera-Garcia, D.
 Mapping Ricci-based theories of gravity into general relativity. 
\emph{Phys. Rev. D} \textbf{2018}, \emph{97}, 021503,
doi:10.1103/PhysRevD.97.021503.


\bibitem{Olmo:2015axa}
Olmo, G.J.; Rubiera-Garcia, D.
 Nonsingular Black Holes in $f(R)$ Theories. 
\emph{Universe} \textbf{2015}, \emph{1}, 173--185,
doi:10.3390/universe1020173.
\bibitem{Olmo:2012nx}
Olmo, G.J.; Rubiera-Garcia, D.
 Reissner-Nordstr\textbackslash{}'om black holes in extended Palatini theories. 
\emph{Phys. Rev. D} \textbf{2012}, \emph{86}, 044014,
doi:10.1103/PhysRevD.86.044014.
\bibitem{Menchon:2017qed}
Menchon, C.; Olmo, G.J.; Rubiera-Garcia, D.
 Nonsingular black holes, wormholes, and de Sitter cores from anisotropic fluids. 
\emph{Phys. Rev. D} \textbf{2017}, \emph{96}, 104028,
doi:10.1103/PhysRevD.96.104028.

\bibitem{Bambi:2015zch}
Bambi, C.; Cardenas-Avendano, A.; Olmo, G.J.; Rubiera-Garcia, D.
 Wormholes and nonsingular spacetimes in Palatini $f(R)$ gravity. 
\emph{Phys. Rev. D} \textbf{2016}, \emph{93}, 064016,
doi:10.1103/PhysRevD.93.064016.
\bibitem{Capozziello:2012hr}
Capozziello, S.; Harko, T.; Koivisto, T.S.; Lobo,~F.S.N.; Olmo, G.J.
 Wormholes supported by hybrid metric-Palatini gravity. 
\emph{Phys. Rev. D} \textbf{2012}, \emph{86}, 127504,
doi:10.1103/PhysRevD.86.127504.


\bibitem{BeltranJimenez:2017doy}
Jimenez, J.B.; Heisenberg, L.; Olmo, G.J.; Rubiera-Garcia, D.
 Born\textendash{}Infeld inspired modifications of gravity. 
\emph{Phys. Rept.} \textbf{2018}, 727, 1--129,
doi:10.1016/j.physrep.2017.11.001.

\bibitem{Kozak:2018vlp}
Kozak, A.; Borowiec, A.
 Palatini frames in scalar\textendash{}tensor theories of gravity. 
\emph{Eur. Phys. J. C} \textbf{2019}, \emph{79}, 335,
doi:10.1140/epjc/s10052-019-6836-y.




\bibitem{reviews}
Liddle, A.R.; Lyth, D.H.  {\it Cosmological Inflation and Large-Scale Structure}; Cambridge University Press: Cambridge, UK, 2000. 
\bibitem{reviews2}
Dodelson, S. {\it Modern Cosmology}; Academic Press: USA 1999.
\bibitem{reviews3}
Mukhanov, V.F. {\it Physical Foundations of Cosmology}; Cambridge University Press: Cambridge, UK, 2005.


\bibitem{Mukhanov:1990me} 
  Mukhanov, V.F.; Feldman, H.A.; Brandenberger, R.H.
   Theory of cosmological perturbations. Part 1. Classical perturbations. Part 2. Quantum theory of perturbations. Part 3. Extensions. 
  \emph{Phys. Rep.} {\bf 1992}, \emph{215}, 203;
\bibitem{Liddle:1999mq} 
  Liddle, A.R.
   An Introduction to cosmological inflation. 
  astro-ph/9901124;
\bibitem{Langlois:2010xc} 
  Langlois, D.
   Lectures on inflation and cosmological perturbations. 
  \emph{Lect. Notes Phys. } {\bf 2010}, \emph{800}, 1.
  
  
 \bibitem{Lidsey:1995np}
Lidsey, J.E.; Liddle, A.R.; Kolb, E.W.; Copeland, E.J.; Barreiro, T.; Abney, M.
 Reconstructing the inflation potential : An overview. 
\emph{Rev. Mod. Phys.} \textbf{1997}, \emph{69}, 373;
 \bibitem{Elizalde:2008yf}
  Elizalde, E.; Nojiri, S.; Odintsov, S.D.; Saez-Gomez, D.; Faraoni, V.
   Reconstructing the universe history, from inflation to acceleration, with phantom and canonical scalar fields. 
  \emph{Phys. Rev. D} {\bf 2008}, \emph{77}, 106005.
 
\bibitem{Ade:2015lrj}
Ade, P.A.R.; Aghanim N.; Arnaud, M.; Arroja F.; Ashdown M.; Aumont J.; Baccigalupi C.; Ballardini M.; Banday A. J.; Barreiro R. B.; et al. [Planck],
 Planck 2015 results. XX. Constraints on inflation. 
\emph{Astron. Astrophys.} \textbf{2016}, \emph{594}, A20;
\bibitem{Akrami:2018odb}
Akrami, Y.; Arroja F.; Ashdown M.; Aumont J.; Baccigalupi C.; Ballardini M.; Banday A. J.; Barreiro R. B.; Bartolo N.; Basak S.; et al. [Planck],
 Planck 2018 results. X. Constraints on inflation. \emph{arXiv} \textbf{2018},  arXiv:1807.06211 [astro-ph.CO].
 
  
  \bibitem{Cognola:2007zu}
  Cognola, G.; Elizalde, E.; Nojiri, S.; Odintsov, S.D.; Sebastiani, L.; Zerbini, S.  A Class of viable modified f(R) gravities describing inflation and the onset of accelerated expansion. 
  \emph{Phys. Rev. D} {\bf 2008}, \emph{77}, 046009;
\bibitem{Nojiri:2003ft}
Nojiri, S.; Odintsov, S.D.
 Modified gravity with negative and positive powers of the curvature: Unification of the inflation and of the cosmic acceleration. 
\emph{Phys. Rev. D} \textbf{2003}, \emph{68}, 123512,
doi:10.1103/PhysRevD.68.123512; 
\bibitem{unifying}
  Nojiri, S.; Odintsov, S.D.    Modified f(R) gravity unifying R**m inflation with Lambda CDM epoch. 
  \emph{Phys. Rev. D} {\bf 2008}, \emph{77}, 026007;
  \bibitem{Cognola:2008zp}
 Cognola,~G.; Elizalde, E.; Odintsov, S.D.; Tretyakov, P.; Zerbini, S.
   Initial and final de Sitter universes from modified f(R) gravity. 
   \emph{Phys. Rev. D} {\bf 2009}, \emph{79}, 044001;
  \bibitem{BambaOdiDie}
  Bamba, K.; Nojiri,~S.; Odintsov,~S.D.; S\'aez-G\'omez, D.
   Inflationary universe from perfect fluid and $F(R)$ gravity and its comparison with observational data. 
   \emph{Phys. Rev. D} {\bf 2014}, \emph{90}, 124061;
    \bibitem{delaCruz-Dombriz:2016bjj}
  Cruz-Dombriz, A.d.; Elizalde, E.; Odintsov,~S.D.; S\'aez-G\'omez, D.
   Spotting deviations from R$^2$ inflation. 
  \emph{JCAP} {\bf 2016}, \emph{1605}, 60;
 \bibitem{Odintsov:2017qif} 
  Odintsov, S.D.; G\'omez, D.S\.; Sharov, G.S.
   Is exponential gravity a viable description for the whole cosmological history?. 
  \emph{Eur. Phys. J. C} {\bf 2017},  \emph{77}, 862;
\bibitem{Sebastiani:2013eqa}
Sebastiani, L.; Cognola, G.; Myrzakulov, R.; Odintsov, S.D.; Zerbini, S.
 Nearly Starobinsky inflation from modified gravity. 
\emph{Phys. Rev. D} \textbf{2014}, \emph{89}, 023518,
doi:10.1103/PhysRevD.89.023518.
\bibitem{Sebastiani:2015kfa}
Sebastiani, L.; Myrzakulov, R.
 F(R) gravity and inflation. 
\emph{Int. J. Geom. Meth. Mod. Phys.} \textbf{2015}, \emph{12}, 1530003,
doi:10.1142/S0219887815300032.
\bibitem{Myrzakulov:2014hca}
Myrzakulov, R.; Odintsov, S.; Sebastiani, L.
 Inflationary universe from higher-derivative quantum gravity. 
\emph{Phys. Rev. D} \textbf{2015}, \emph{91}, 083529,
doi:10.1103/PhysRevD.91.083529. 
\bibitem{Bamba:2014jia}
Bamba, K.; Myrzakulov, R.; Odintsov, S.D.; Sebastiani, L.
 Trace-anomaly driven inflation in modified gravity and the BICEP2 result. 
\emph{Phys. Rev. D} \textbf{2014}, \emph{90}, 043505,
doi:10.1103/PhysRevD.90.043505.
  
  
\bibitem{Starobinsky:1980te}
 Starobinsky, A.A.
   A New Type of Isotropic Cosmological Models Without Singularity. 
  \emph{Phys.  Lett.}  {\bf 1980}, \emph{91B},~99.
\newpage


\bibitem{Shimada:2018lnm}
Bauer, F.; Demir, D.A.
 Inflation with Non-Minimal Coupling: Metric versus Palatini Formulations. 
\emph{Phys. Lett. B} \textbf{2008}, \emph{665}, 222--226,
doi:10.1016/j.physletb.2008.06.014; 
\bibitem{ShiMaeda}
Shimada, K.; Aoki, K.; Maeda, K.I.
 Metric-affine Gravity and Inflation. 
\emph{Phys. Rev. D} \textbf{2019}, \emph{99}, 104020,
doi:10.1103/PhysRevD.99.104020; 
\bibitem{Gialamas:2020snr}
Gialamas, I.D.; Karam, A.; Racioppi, A.
 Dynamically induced Planck scale and inflation in the Palatini formulation. 
\bibitem{Tenkanen:2020dge}
Tenkanen,~T. 
 Tracing the high energy theory of gravity: an introduction to Palatini inflation. 
\emph{Gen. Rel. Grav.} \textbf{2020}, \emph{52}, 33,
doi:10.1007/s10714-020-02682-2; 
\bibitem{Das:2020kff}
Das, N.; Panda, S.
 Inflation in f(R,h) theory formulated in the Palatini formalism. 
\bibitem{Jarv:2017azx}
J\"arv, L.; Racioppi, A.; Tenkanen, T.
 Palatini side of inflationary attractors. 
\emph{Phys. Rev. D} \textbf{2018}, \emph{97}, 083513,
doi:10.1103/PhysRevD.97.083513. 

\bibitem{Tenkanen:2017jih}
Tenkanen, T.
 Resurrecting Quadratic Inflation with a non-minimal coupling to gravity. 
\emph{JCAP} \textbf{2017}, \emph{12}, 1,
doi:10.1088/1475-7516/2017/12/001. 
\bibitem{Markkanen:2017tun}
Markkanen, T.; Tenkanen, T.; Vaskonen, V.; Veerm\"ae, H.
 Quantum corrections to quartic inflation with a non-minimal coupling: metric vs. Palatini. 
\emph{JCAP} \textbf{2018}, \emph{3}, 29,
doi:10.1088/1475-7516/2018/03/029. 
\bibitem{Antoniadis:2018yfq}
Antoniadis, I.; Karam, A.; Lykkas,~A.; Pappas, T.; Tamvakis, K.
 Rescuing Quartic and Natural Inflation in the Palatini Formalism. 
\emph{JCAP} \textbf{2019}, \emph{3}, 5.
doi:10.1088/1475-7516/2019/03/005. 


\bibitem{Antoniadis:2018ywb}
Antoniadis, I.; Karam, A.; Lykkas, A.; Tamvakis, K.
 Palatini inflation in models with an $R^2$ term. 
\emph{JCAP} \textbf{2018}, \emph{11}, 28,
doi:10.1088/1475-7516/2018/11/028. 
\bibitem{Enckell:2018hmo}
Enckell, V.M.; Enqvist, K.; Rasanen, S.; Wahlman, L.P.
 Inflation with $R^2$ term in the Palatini formalism. 
\emph{JCAP} \textbf{2019}, \emph{2}, 22,
doi:10.1088/1475-7516/2019/02/022. 
\bibitem{Edery:2019txq}
Edery, A.; Nakayama, Y.
 Palatini formulation of pure $R^2$ gravity yields Einstein gravity with no massless scalar. 
\emph{Phys. Rev. D} \textbf{2019}, \emph{99}, 124018,
doi:10.1103/PhysRevD.99.124018. 
\bibitem{Stachowski:2016zio}
Stachowski, A.; Szyd\l{}owski,~M.; Borowiec, A.
 Starobinsky cosmological model in Palatini formalism. 
\emph{Eur. Phys. J. C} \textbf{2017}, \emph{77}, 406,
doi:10.1140/epjc/s10052-017-4981-8. 


\bibitem{Jarv:2020qqm}
J\"arv, L.; Karam, A.; Kozak, A.; Lykkas, A.; Racioppi, A.; Saal, M.
 Equivalence of inflationary models between the metric and Palatini formulation of scalar-tensor theories. 
\emph{Phys. Rev. D} \textbf{2020}, \emph{102}, 044029,
doi:10.1103/PhysRevD.102.044029. 

\bibitem{Koivisto:2005yc}
Koivisto, T.; Kurki-Suonio, H.
 Cosmological perturbations in the palatini formulation of modified gravity. 
\emph{Class. Quant. Grav.} \textbf{2006}, \emph{23}, 2355--2369,
doi:10.1088/0264-9381/23/7/009.
\bibitem{Tamanini:2010uq}
Tamanini, N.; Contaldi, C.R.
 Inflationary Perturbations in Palatini Generalised Gravity. 
\emph{Phys. Rev. D} \textbf{2011}, \emph{83}, 044018,
doi:10.1103/PhysRevD.83.044018. 
\bibitem{Fu:2017iqg}
Fu, C.; Wu, P.; Yu, H.
 Inflationary dynamics and preheating of the nonminimally coupled inflaton field in the metric and Palatini formalisms. 
\emph{Phys. Rev. D} \textbf{2017}, \emph{96}, 103542,
doi:10.1103/PhysRevD.96.103542. 


\bibitem{Gialamas:2019nly}
Gialamas, I.D.; Lahanas, A.B.
 Reheating in $R^2$ Palatini inflationary models. 
\emph{Phys. Rev. D} \textbf{2020}, \emph{101}, 084007,
doi:10.1103/PhysRevD.101.084007. 
\bibitem{Rubio:2019ypq}
Rubio, J.; Tomberg, E.S.
 Preheating in Palatini Higgs inflation. 
\emph{JCAP} \textbf{2019}, \emph{4}, 21,
doi:10.1088/1475-7516/2019/04/021. 

\bibitem{Rasanen:2018ihz}
Bauer, F.; Demir, D.A.
 Higgs-Palatini Inflation and Unitarity. 
\emph{Phys. Lett. B} \textbf{2011}, \emph{698}, 425--429,
doi:10.1016/j.physletb.2011.03.042. 
\bibitem{Rasanen:2018Gu}
Rasanen, S.
 Higgs inflation in the Palatini formulation with kinetic terms for the metric. 
\emph{Open J. Astrophys.} \textbf{2019}, \emph{2}, 1,
doi:10.21105/astro.1811.09514. 
\bibitem{Gialamas:2020vto}
Gialamas, I.D.; Karam, A.; Lykkas, A.; Pappas, T.D.
 Palatini-Higgs inflation with nonminimal derivative coupling. 
\emph{Phys. Rev. D} \textbf{2020}, \emph{102}, 063522,
doi:10.1103/PhysRevD.102.063522. 

\bibitem{Racioppi:2017spw}
Racioppi, A.
 Coleman-Weinberg linear inflation: metric vs. Palatini formulation. 
\emph{JCAP} \textbf{2017}, \emph{12}, 41,
doi:10.1088/1475-7516/2017/12/041. 

\end{thebibliography}
\end{document}